\documentclass[11pt,twocolumn]{article}
\usepackage[utf8]{inputenc}
\usepackage[version=4]{mhchem}
\usepackage{multirow}
\usepackage{graphicx}
\usepackage{physics}
\usepackage{amssymb}
\usepackage{textgreek}
\usepackage{gensymb}
\usepackage{float}
\usepackage[]{caption}
\usepackage[margin=0.7in]{geometry}
\usepackage{titletoc}

\usepackage{calrsfs}
\newcommand{\R}{\mathcal{R}}

\usepackage{hyperref}
\hypersetup{
    colorlinks = true,
    allcolors=black,
}

\usepackage[
   backend=bibtex,
  citestyle=numeric-comp,
  style=chem-rsc,
    maxbibnames=1,
    sorting=none
]{biblatex}
\usepackage{makecell} 

\usepackage[labelfont=bf]{caption}
\usepackage{abstract}

\usepackage{titlesec}
\titlespacing\section{0pt}{12pt plus 4pt minus 2pt}{8pt plus 4pt minus 2pt}
\titleformat{\section}[block]{\scshape\filcenter\bfseries\large}{}{0.8em}{}
\titleformat{\subsection}[block]{\scshape\filcenter}{}{1em}{}

\usepackage{upgreek}
\usepackage{mathptmx}

\newcommand\taba[1][0.14cm]{\hspace*{#1}}
\usepackage{changepage}
\usepackage{physics}

\begin{document}

\twocolumn[{
    \begin{flushleft}
        \Large\textbf{Universal conductance fluctuations in a \ce{MnBi2Te4} thin film}
        
    \end{flushleft}
    
    \begin{flushleft}
        {Molly P. Andersen$^{1,2}$, Evgeny Mikheev$^{3,2,4}$, Ilan T. Rosen$^{5,2,6}$, Lixuan Tai$^7$, Peng Zhang$^7$, Kang L. Wang$^7$, Marc A. Kastner$^{3,2,8}$, and David Goldhaber-Gordon$^{3,2,a}$}
    \end{flushleft}
    
    \begin{flushleft}
        \footnotesize{$^1$\textit{Department of Materials Science and Engineering, Stanford University, Stanford, California 94305, USA}}\\
        \footnotesize{$^2$\textit{Stanford Institute for Materials and Energy Sciences, SLAC National Accelerator Laboratory, 2575 Sand Hill Road, Menlo Park, California 94025, USA}}\\
        \footnotesize{$^3$\textit{Department of Physics, Stanford University, Stanford, California 94305, USA}}\\
        \footnotesize{$^4$\textit{Department of Physics, University of Cincinnati, Cincinnati, Ohio 45221, USA}}\\
        \footnotesize{$^5$\textit{Department of Applied Physics, Stanford University, Stanford, California 94305, USA}}\\
        \footnotesize{$^6$\textit{Research Laboratory of Electronics, Massachusetts Institute of Technology, Cambridge, MA, 02139}}\\
        \footnotesize{$^7$\textit{Department of Electrical and Computer Engineering, Department of Physics and Astronomy, University of California, Los Angeles, California 90095, USA}}\\    
        \footnotesize{$^8$\textit{Department of Physics, Massachusetts Institute of Technology, Cambridge, Massachusetts 02139, USA}}\\    
        \footnotesize{$^a$To whom correspondence should be addressed; E-mail: \texttt{goldhaber-gordon@stanford.edu}}
    \end{flushleft}   

\begin{abstract}

Quantum coherence of electrons can produce striking behaviors in mesoscopic conductors, including weak localization and the Aharonov-Bohm effect. Although magnetic order can also strongly affect transport, the combination of coherence and magnetic order has been largely unexplored. Here, we examine quantum coherence-driven universal conductance fluctuations in the antiferromagnetic, canted antiferromagnetic, and ferromagnetic phases of a thin film of the topological material \ce{MnBi2Te4}. In each magnetic phase we extract a charge carrier phase coherence length of about 100~nm. The conductance magnetofingerprint is repeatable when sweeping applied magnetic field within one magnetic phase, but changes when the applied magnetic field crosses the antiferromagnetic/canted antiferromagnetic magnetic phase boundary. Surprisingly, in the antiferromagnetic and canted antiferromagnetic phase, but not in the ferromagnetic phase, the magnetofingerprint depends on the direction of the field sweep. To explain these observations, we suggest that conductance fluctuation measurements are sensitive to the motion and nucleation of magnetic domain walls in \ce{MnBi2Te4}. 
\newline
\

\end{abstract}

}]


The van der Waals material \ce{MnBi2Te4} (MBT) has drawn recent excitement based on its topological properties and multiple magnetic phases. Below the N\'eel temperature $\sim20$~K, within each van der Waals layer magnetic moments on the Mn sites order ferromagnetically with an out-of-plane easy axis. Depending on external magnetic field and film thickness a variety of interlayer orderings have been observed, some associated with quantized Hall resistance or axion insulating states~\cite{otrokov2019,li2019,zhang2019,deng2020,liu2020,liu2021,ovchinnikov2021,lin2022,ovchinnikov2022}. The well-defined stoichiometry implied by the chemical formula suggests that this should be a cleaner system than the more widely studied quaternary alloy \ce{Cr_x(Bi,Sb)_{2-x}Te3}~\cite{chang2013,mogi2015,fox2018}. However, in practice disorder is ubiquitous in both thin-film and bulk single-crystal samples of MBT. Sources of defects include antisite defects~\cite{zeugner2019,huang2020,garnica2022}, impurity phases~\cite{hou2020,zhu2020,zhao2021,tai2022}, and magnetic domains~\cite{sass2020,sass2020b}. These have hindered experimental validation of striking predictions for behavior near $B=0$ such as quantization of Hall resistance and persistence of topological phenomena at elevated temperatures~\cite{chen2019}.

Regardless of a system's topology, electrical transport measurements are sensitive to charge carrier scattering through the material's disorder landscape. When device features become similar in size or smaller than the phase coherence length, wavelike interference between different trajectories can have dramatic effects on electronic transport measurements~\cite{anderson1979,ab1959}. Magnetism is known to suppress charge carrier coherence through scattering off of magnetic disorder, domain walls, or spin waves~\cite{chandrasekhar1990,takane2003,koyama2003}. Few works have explored residual coherence-driven phenomena in magnetically ordered materials~\cite{aumentado2000,kasai2003,lee2004,wagner2006,lee2007,vila2007,neumaier2007}. Fewer still have explored charge carrier coherence within and between different magnetic phases of the same material. 

In this work, we study low-temperature electronic transport through Hall bars patterned from a thin film of MBT grown by molecular beam epitaxy~\cite{tai2022}. We observe Universal Conductance Fluctuations (UCFs)~\cite{lee1985}, a familiar manifestation of coherent transport through a disordered material, in all three magnetic phases of MBT. We analyze the UCFs as a function of device geometry, magnetic field, and temperature to gain insight into charge carrier scattering mechanisms, phase coherence lengths, and timescales for rearrangement of the elastic scatterer disorder landscape. Surprisingly, our UCF data appear sensitive to the presence and motion of magnetic domains in MBT's anti-ferromagnetic (AFM) and canted anti-ferromagnetic (cAFM) phases, as well as the nucleation of new domain patterns at the AFM/cAFM magnetic phase transition. We note that we do not observe quantized resistivity in our devices, implying that neither Chern nor axion insulating states are established, presumably due to doping or disorder~\cite{zhu2020,garrity2021,tai2022,liu2022}.


\begin{figure}[h]
\centering
	\includegraphics[width=0.3\textwidth]{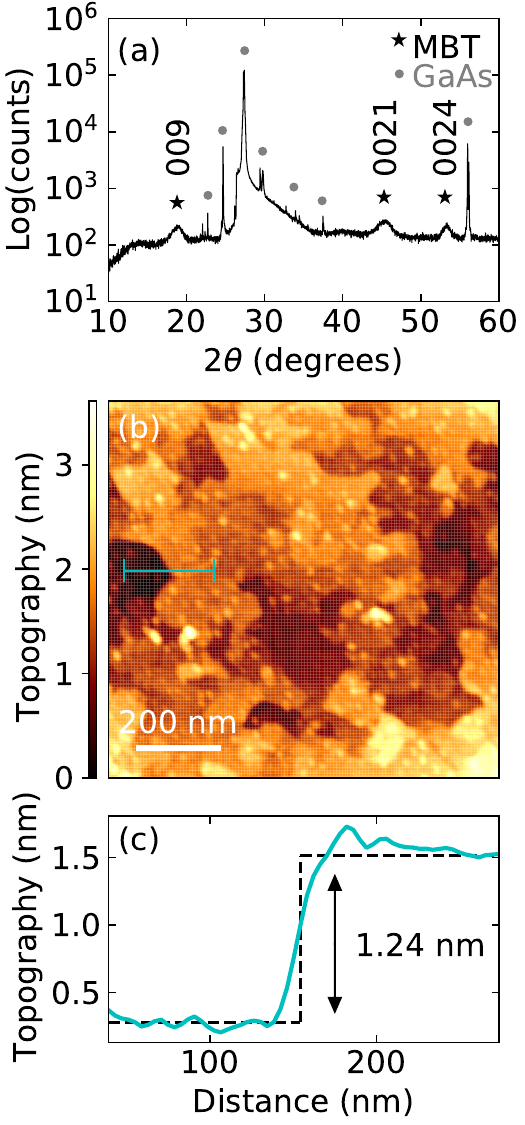}
	\caption{(a) The x-ray diffraction pattern of the 6 SL MBT thin film on a GaAs substrate. Peaks associated with MBT (GaAs substrate~\cite{tai2022}) are indicated with black stars (grey circles). (b) Topography of the MBT thin film, acquired with atomic force microscopy. Scale bar, 200 nm. (c) Line cut of topography data along the cyan line in (b). Data were averaged across 15 parallel traces spanning 60 nm in the transverse direction, indicated by the extent of the crosspieces appended to the tails of the cyan line.}
	\label{fig1}
\end{figure}

The \ce{MnBi2Te4} film used in this work~\cite{tai2022} was approximately 6 van der Waals layers (called septuple layers (SLs)) thick and grown on a semi-insulating GaAs (111)B substrate via molecular beam epitaxy~\cite{sup}. X-ray diffraction was used to characterize the film's crystal structure (Figure~\ref{fig1}(a)). Three (00l) (l = 9, 21, 24) reflections from the MBT film are apparent, as are two high-intensity (111) and (222) peaks from the GaAs substrate, confirming epitaxial growth with (001)R of the MBT film parallel to (111)C of the GaAs substrate. From the (00l) XRD reflections, the out-of-plane lattice constant of MBT is $c=40.876 \pm 0.003$~\AA, in agreement with other reports of MBT lattice parameters~\cite{yan2019,zeugner2019,zhu2020,bac2022}. Since one unit cell of MBT is composed of 3 SLs, our data indicate 1~SL = 13.63~\AA. Topography of the film, as measured by atomic force microscopy several months after film growth, is shown in Figure~\ref{fig1}(b). A linecut across a step edge (Figure~\ref{fig1}(c)) yields a height 1.24~nm, reasonably consistent with a single SL step. Steps with smaller heights are also visible in Figure~\ref{fig1}(b); these may reflect phases besides the \ce{MnBi2Te4} stoichiometry produced during growth or through surface reconstruction~\cite{hou2020,zhu2020,zhao2021,tai2022}, or may result from post-growth sample aging.

Two standard Hall bar devices were fabricated on a single chip of this film. Throughout device fabrication, care was taken to ensure minimal thermal or chemical degradation of the sample~\cite{sup}. Each Hall bar featured two pairs of voltage taps separated by two squares of material; in Device A (B), one square was $10 \times 10$ ($20 \times 20$)~$\mu$m$^2$. These were larger than measured electronic phase coherence lengths in the material, as discussed below. Optical images of both devices are shown in the supplemental materials. 

\begin{figure*}[h]
\centering
	\includegraphics[width=\textwidth]{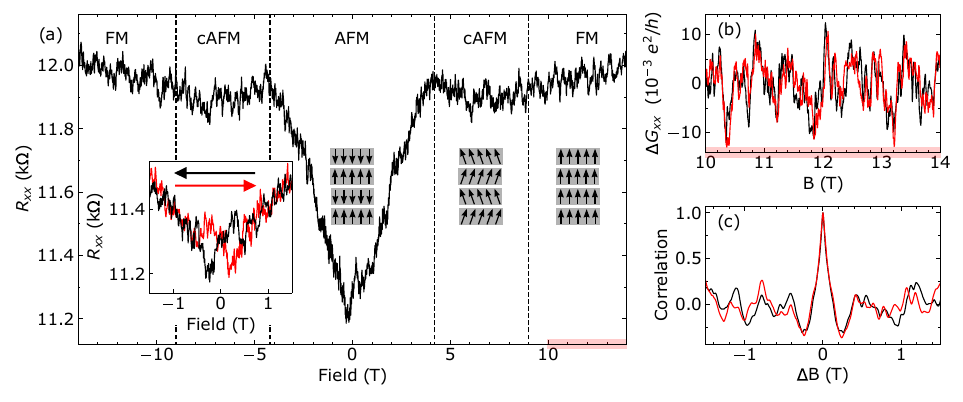}
	\caption{(a) Longitudinal resistance $R_{xx}$ as a function of out-of-plane applied magnetic field, swept from positive to negative values. Field values at which magnetoresistance qualitatively changes, corresponding to magnetic phase transitions identified in previous works, are marked by dashed vertical lines. The corresponding orientation of Mn magnetic moments within successive layers of MBT (grey boxes) is indicated by black arrows. Inset: $R_{xx}$ near zero applied field for each field sweep direction.  (b) Conductance fluctuations $\Delta G_{xx}$ as a function of magnetic field for $B=10-14$ T, extracted by subtracting a polynomial background from the same measurements shown in (a) and from a corresponding upward field sweep. The red box on the field axis in (a) indicates the subset of data plotted in (b). (c) Correlation of $\Delta G_{xx}(B)$ and $\Delta G_{xx}(B+\Delta B)$ as a function of $\Delta B$, calculated for the data shown in (b) acquired between 10--14~T. For all plots, red (black) indicates field sweeping up (down).}
	\label{fig2}
\end{figure*}

Electronic transport measurements were taken on Device A and Device B simultaneously in a dilution refrigerator with a base temperature of 35~mK using 5--8~Hz few-nA current excitations and standard lock-in techniques. Data from the top longitudinal voltage pair of Device A are presented in the main text. Data from Device B, presented in the supplemental materials, show no substantial differences from those from Device A beyond an expected geometric scaling. Although devices feature electrostatic top gates, for all measurements presented in the main text the electrostatic top gates were explicitly grounded to avoid time-dependent conductance fluctuations associated with changes in charge trap population as and after the gate voltage is swept~\cite{sup}. Where measurements were acquired as a function of applied external magnetic fields, the field was incremented in either 2 or 5~mT steps. We term positive field values to be parallel to the upward-facing sample normal. Full details of device fabrication, measurement techniques, and measurement parameters are presented in the supplemental materials. 


Figure~\ref{fig2}(a) plots longitudinal resistance $R_{xx}=V_{xx}/I$~\cite{sup} as a function of applied magnetic field at base temperature of about 35 mK. Below the N\'eel temperature of about 20 K~\cite{su2021}, MBT orders into an out-of-plane antiferromagnetic (AFM) phase for weak externally applied magnetic field. The spin arrangement is schematically illustrated within the figure for this and other phases that arise as the external magnetic field is increased: a canted antiferromagnet (cAFM), followed by ferromagnetic (FM) interlayer coupling at high fields~\cite{ovchinnikov2021}. Though our electronic transport measurements give no direct insight into the exact magnetic ordering at any applied field, vertical dashed lines mark qualitative changes in magnetoresistance that agree with magnetic phase boundaries observed in other studies~\cite{liu2020,ovchinnikov2021,tai2022,zhao2021,bac2022}. Hall measurements acquired simultaneously reveal a charge carrier density of about $4\times 10^{12}$ cm$^{-2}$ electrons, and permit analysis of the charge carrier mobility as a function of applied field (Figure~S3)~\cite{sup}.

Background magnetoresistance is roughly independent of field sweep direction at fields $>\abs{1}$ T. Across 0~T, however, a hysteretic dip in magnetoresistance occurs (Figure~\ref{fig2}(a) left inset). This low-field butterfly hysteresis could be caused by \ce{MnTe_x} or Mn-doped \ce{Bi2Te3} magnetic impurities that often condense with MBT during growth or form upon air exposure and aging~\cite{hou2020,zhu2020,zhao2021,tai2022}, although the XRD data from our film present no evidence of these phases. Other possible explanations for the butterfly hysteresis include flipping of uncompensated magnetic moments in odd-layer regions of the MBT film or surface spin flop transitions which have been observed to occur at low fields, prior to the bulk spin flop transition~\cite{sass2020b}. Regardless of its origin, this butterfly hysteresis prevents a standard analysis of (anti-)localization magnetoresistance peaks near $B=0$~T.

We focus instead on fluctuations in $R_{xx}$ apparent in Figure~\ref{fig2}(a) across the entire magnetic field range studied. As discussed below, these fluctuations were highly reproducible upon sweeping field up and down over several Tesla (Figure~\ref{fig2}(b)) but apparently random (uncorrelated beyond a small field scale, Figure~\ref{fig2}(c)). The phenomenology of these fluctuations is consistent with universal conductance fluctuations, motivating the analysis below, and is inconsistent with other known mechanisms of quantum oscillations (Section~S4)~\cite{sup}.


As a charge carrier propagates, wavelike constructive and destructive interference between different paths affects transmission probabilities, modulating the local conductance of the sample. In macroscopic structures, the charge carrier phase coherence length $\ell_{\phi}$ is far smaller than sample dimensions, so quantum coherence-driven conductance changes average out. However, when $\ell_{\phi}$ exceeds a sample's dimensions measured conductance depends on details of the disorder landscape, and exhibits fluctuations as a function of any control parameter that adjusts the relative phases of different paths. These conductance fluctuations have a characteristic scale of $e^2/h$ independent of sample details (hence the term ``universal''), where $e$ is electron charge and $h$ is Plank's constant~\cite{lee1985,beenaker1991}. Magnetic field modulates the relative phase of different paths, so a hallmark feature of UCF is conductance variations that look like random noise but are in fact a consistent function of applied magnetic field so long as the disorder landscape remains stable.

Figure~\ref{fig2}(b) shows $\Delta G_{xx}$ -- fluctuations in conductance $G_{xx} = 1/R_{xx}$ after subtracting a best-fit fourth-order polynomial to remove smooth background magnetoresistance. These conductance fluctuations within the high-field FM phase have root-mean-squared magnitude $\Delta G_\mathrm{rms} = 0.0045~e^2/h$. Measurements of $\Delta G_\mathrm{rms} \ll e^2/h$ are well-established in UCF literature~\cite{imry1986,skocpol1986}: when $\ell_{\phi}$ is smaller than the device dimensions, transport is coherent within local portions of the device but incoherent averaging between regions separated by more than $\ell_{\phi}$ reduces the measured $\Delta G_\mathrm{rms}$. For our device dimensions of width $w=10~\mu$m and length $l=20~\mu$m, our observation of 200-fold reduction of $\Delta G_\mathrm{rms}$ compared to a fully coherent scenario suggests $\ell_{\phi}\approx (l,w)/200 \approx 100$~nm. 

A thermal spread of charge carrier energy and thus momentum can also suppress $\Delta G_\mathrm{rms}$, since carriers of different momentum accumulate different phase along the same path~\cite{lee1985,lee1987,beenaker1991}. Based on our measured conductivity and estimated Fermi velocity we calculate that phase is thermally randomized over a length $\ell_T\approx2~\mu$m (Section~S3~\cite{sup}). Since $\ell_T \gg \ell_{\phi}$, thermal smearing can be neglected at the electron temperatures attained in our experiments (discussed below and in Section~S6~\cite{sup}).  

Where thermal averaging can be neglected, \[\Delta G_\mathrm{rms} =\frac{e^2}{h} c_d \sqrt{\frac{ks^2}{\beta}} \sqrt{\frac{\ell_{\phi}^2}{wl}}.\] Constants $c_d$ and $\beta$ depend on sample dimensionality ($c_d=0.862$ for two dimensions~\cite{lee1985}) and symmetries of the ensemble of disordered wavefunctions ($\beta=2$ when time-reversal symmetry is broken~\cite{lee1987}), respectively. Parameters $s$ and $k$ account for multiplicity of spin and other degeneracies. These degeneracies are challenging to determine for the MBT thin film used in this work. Even uniform MBT has a complicated band structure--featuring multiple topological surface states in addition to topologically trivial bulk states~\cite{li2019,chen2019}--and sample-to-sample variations~\cite{otrokov2019,zeugner2019,chen2019,estyunin2020,garnica2022} only complicate the picture. In our disordered film with varying layer number and no specific knowledge of the location of the Fermi level relative to band structure features, we use $k=s=1$ and acknowledge that the absolute magnitude of $\ell_{\phi}$ we determine may accordingly be off by a constant multiple. With these constants, we find $\ell_{\phi} =104$~nm in the FM phase.

The Pearson product-moment correlation coefficient $\R$ quantifies the degree of correlation between a pair of field sweeps $i, j$. Calculated as \[\R = \frac{\sum (\Delta G_{i}(B) - \overline{\Delta G_{i}})(\Delta G_{j}(B) - \overline{\Delta G_{j}})}{\sqrt{\sum (\Delta G_{i}(B) - \overline{\Delta G_{i}})^2(\Delta G_{j}(B) - \overline{\Delta G_{j}})^2}},\] $\R=1$ for a pair of perfectly correlated sweeps and $\R=0$ for a pair of perfectly uncorrelated sweeps. For the pair of field sweeps shown in Figure~\ref{fig2}(b), $\R=$ 0.73, in accord with the strong visible similarity, despite the $\sim25$~hour separation between the acquisition of the two field sweeps. A discussion of repeatability of measured UCFs over varied time scales is presented in Section~S5~\cite{sup}.

Correlation of UCFs as a function of separation in applied field rather than in time provides direct insight into $\ell_{\phi}$~\cite{lee1985,skocpol1986}. Changing field by an amount sufficient to thread a single flux quantum through a single coherent area largely randomizes the relative phases of different trajectories. This ``coherence field'' scale $B_c$ is therefore related to $\ell_{\phi}^{-2}$ and can be extracted as the half-width at half-max of the correlation function $C(\Delta B) = \langle \Delta G_{xx}(B) \Delta G_{xx}(B+\Delta B) \rangle$. In two dimensions, \[B_c = 0.49 \frac{h}{e} \frac{1}{\ell_{\phi}^2},\] provided that thermal averaging can be neglected~\cite{lee1985,lundeberg2012}. Figure~\ref{fig2}(c) shows $C(\Delta B)$ for the $\Delta G_{xx}$ traces plotted in Figure~\ref{fig2}(b). The corresponding $B_c$ is 40~mT, yielding $\ell_{\phi} \approx 240$~nm, within a factor of 3 of the estimate obtained via $\Delta G_\mathrm{rms}$. Given the uncertainty associated with $B_c$ as a metric for phase coherence in two dimensions~\cite{lundeberg2012}, the two methods produce acceptable agreement. Due to the limited field domains over which most data were acquired, further discussion of $B_c$ is confined to the supplemental materials. Additional estimates of $\ell_{\phi}$ below instead use the $\Delta G_\mathrm{rms}$ method.

Since MBT is a magnetic material, one might question whether its magnetic disorder has an impact on UCF measurements. Magnetic force microscopy measurements on single-crystal MBT have revealed spatial domains of magnetization that evolve with magnetic field within each magnetic phase and across phase boundaries~\cite{sass2020}. To study the impact of magnetic disorder on our UCFs, we next present measurements of $\Delta G_{xx}$ upon repeated sweeps within each magnetic phase. Then we examine how this changes if we cross a magnetic phase boundary before returning for a subsequent sweep in an initial phase. These UCF measurements appear sensitive to the motion and nucleation of MBT's magnetic domains.

\begin{figure*}[ht]
\centering
	\includegraphics[width=\textwidth]{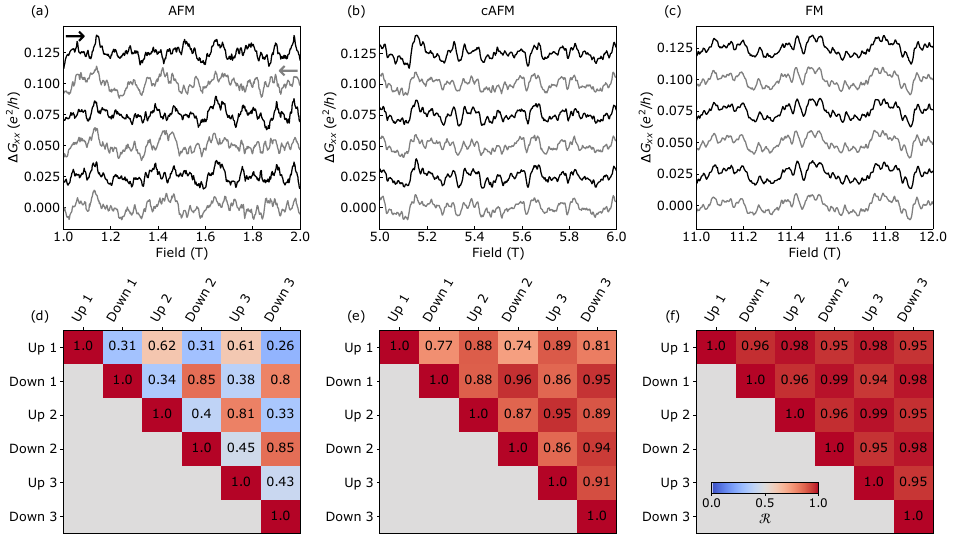}
	\caption{Repeatability of UCF, with attention to field sweep direction. (a-c) $\Delta G_{xx}$ as a function of applied field for sweeps up (black) and down (grey) in the (a) AFM ($\Delta G_\mathrm{rms}=0.0049~e^2/h$; value averaged across all sweeps), (b) cAFM ($\Delta G_\mathrm{rms}=0.0043~e^2/h$), and (c) FM ($\Delta G_\mathrm{rms}=0.0050~e^2/h$) phases. For each magnetic phase, field was swept back and forth across a 1 T range in immediately subsequent runs. (d-f) Correlation $\R$ between each separate pair of $\Delta G_{xx}$ traces within the (d) AFM, (e) cAFM, and (f) FM phases. In the AFM and cAFM phases, runs with the same field sweep direction are markedly more highly correlated than runs with opposite field sweep directions. The slight reduction in correlation between the initial (``Up 1'') sweep and subsequent sweeps in the AFM and cAFM phases can likely be attributed to slight cooling over time; all three datasets were acquired after measurements at elevated temperatures and allowing the system to return to base temperature.}
	\label{fig3}
\end{figure*}

UCFs occur in each magnetic phase of MBT. Figures~\ref{fig3}(a-c) present changes in longitudinal conductance $\Delta G_{xx}$ while applied magnetic field was swept continuously back and forth across a 1~T range. Data acquired while sweeping the field up (down) are shown in black (grey). Repeatable conductance fluctuation patterns indicative of UCF are apparent, particularly in the cAFM and FM phases. In each magnetic phase, $\Delta G_\mathrm{rms} =  0.004-0.005~e^2/h$, indicating $\ell_{\phi} \geq 90$~nm across the entire magnetic phase diagram. The specific magnitude of $\Delta G_\mathrm{rms}$ and therefore $\ell_{\phi}$ is sensitive to the size of the current excitation, likely due to current-driven heating as discussed below and Section~S6~\cite{sup}.

Figures~\ref{fig3}(d-f) plot the correlation between each pair of field sweeps within each magnetic phase. In the FM phase $\R>0.95$ for nearly all pairs of field sweeps. In contrast, in the AFM and cAFM phases the magnitude of $\R$ strongly depends on whether or not magnetic field was swept in the same direction for both $\Delta G_{i}$ and $\Delta G_{j}$; correlation between runs with opposite field sweep direction is clearly reduced. This effect is not typically observed in measurements of UCFs and is likely related to magnetic domain reconfiguration, as discussed below. 


\begin{figure*}[h]
\centering
	\includegraphics[width=17.2cm]{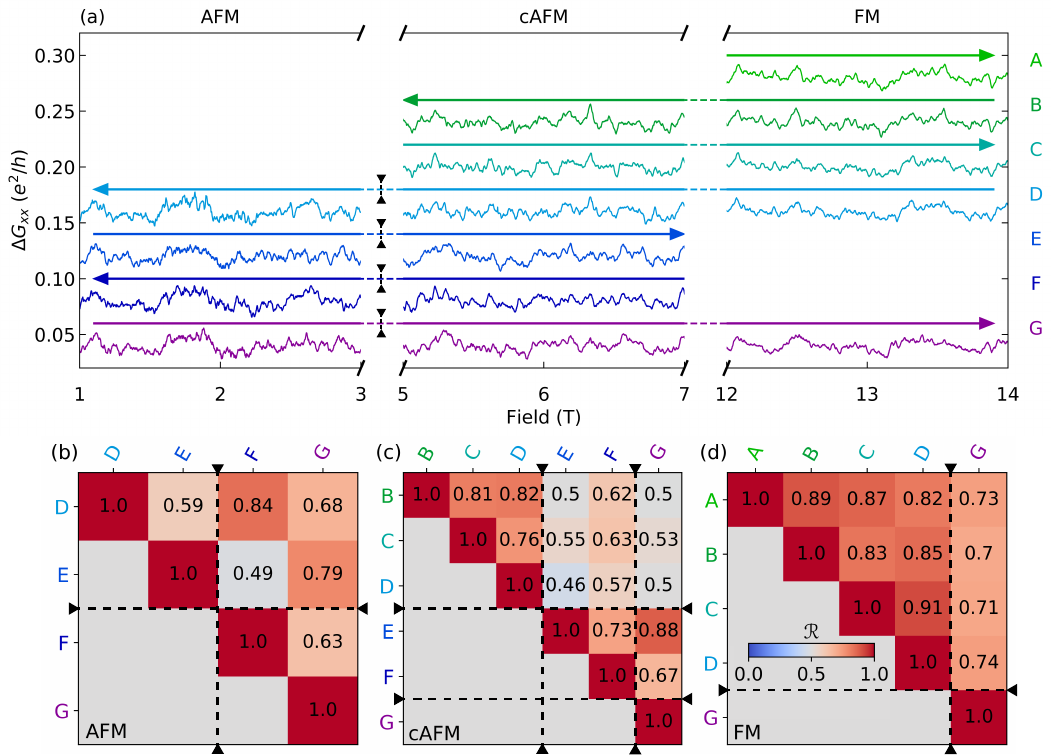}
	\caption{Repeatability of UCF pattern with intervening crossing of magnetic phase boundaries. (a) $\Delta G_{xx}$ as a function of magnetic field. Measurements were taken in order from top to bottom. Each sweep is labelled with a letter on the right side of the plot, and presented in a unique color. Corresponding arrows indicate field sweep direction. (b-d) Repeatability $\R$ of $\Delta G_{xx}$ between field sweeps presented in (a) for the (b) AFM, (c) cAFM, and (d) FM phases. Black dashed lines indicate where the AFM/cAFM phase boundary was crossed.}
	\label{fig4}
\end{figure*}

Figure~\ref{fig4}(a) presents $\Delta G_{xx}$ as magnetic field was swept continuously across specific magnetic phase transitions. Data collection started at 12~T at the top of Figure~\ref{fig4}(a) and proceeded from top to bottom, with the direction of the seven field sweeps (lettered ``A" through ``G") indicated by the corresponding arrows. Field sweep ranges were chosen to isolate behavior in each magnetic phase before and after crossing each magnetic phase boundary. Figure~\ref{fig4}(b-d) shows $\R$ for the pairs of field sweeps acquired within each magnetic phase. In the cAFM and FM phases, we observe that the repeatability between traces is generally lower when the AFM/cAFM phase boundary (at which the magnetic ordering undergoes a spin-flop transition) was crossed between the two sweeps; for example, in the cAFM phase, $\R_\text{D-E}<\R_\text{E-F}$, although in one case this trend is not followed ($\R_\text{E-G}>\R_\text{E-F}$). Below, we attribute this reduced correlation to reconfiguration of magnetic domains.


\begin{figure*}[ht]
\centering
	\includegraphics[width=17cm]{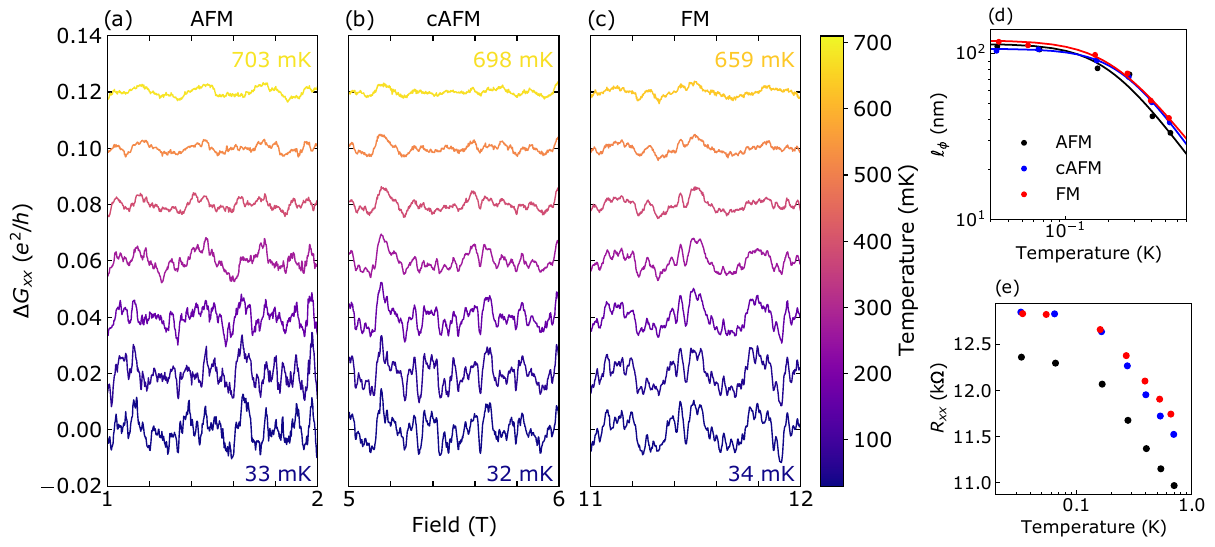}
	\caption{Temperature dependence of UCF. (a-c) $\Delta G_{xx}$ as a function of applied magnetic field in the (a) AFM, (b) cAFM, and (c) FM phases. Temperature is indicated by the color scale, and increases from bottom to top. (d) Phase coherence length $\ell_{\phi}$ as a function of temperature, fit to $\ell_{\phi}^{-2} =  a + bT^p$ with free parameters $a$, $b$, and $p$ in each magnetic phase. For the AFM phase, fitting produces $\ell_{\phi}^{-2} \sim T^{2.4}$. For the cAFM phase, $\ell_{\phi}^{-2} \sim T^{2.3}$. For the FM phase, $\ell_{\phi}^{-2} \sim T^{2.4}$. (e) Longitudinal resistance $R_{xx}$ as a function of temperature extracted from the same measurements as used to generate (a-d). Values of $R_{xx}$ are plotted for external magnetic field 1.5~T (AFM, black), 5.5~T (cAFM, blue) and 11.5~T (FM, red).}
	\label{fig5}
\end{figure*}

To complete our characterization of the phenomenology of UCFs in MBT, we next examine the temperature dependence of $\Delta G_{xx}$. As temperature increases, coherence-breaking scattering mechanisms are activated. The temperature scaling of the UCF-determined $\ell_{\phi}$ can be used to determine the dominant coherence-breaking scattering mechanism. Figure~\ref{fig5}(a-c) shows $\Delta G_{xx}$ traces as a function of applied field at various temperatures for each magnetic phase. Increasing temperature decreased the magnitude of the fluctuations as expected. 

In literature, temperature scaling relationships are often expressed in terms of the dephasing rate $\tau_{\phi}^{-1} = (D\ell_{\phi})^{-2} \propto  T^p$ for temperature $T$, diffusivity $D$, and coherence time $\tau_{\phi}$~\cite{altshuler1980,altshuler1982,lin2002,neumaier2008}. Common scaling relationships include $p=1$, which indicates electron-electron scattering is dominant~\cite{altshuler1980,altshuler1982}, and $p$ between 2 and 4, which indicates electron-phonon scattering is dominant~\cite{lin2002}. Figure~\ref{fig5}(d) plots $\ell_{\phi}$, as determined from $\Delta G_\mathrm{rms}(T)$, as a function of temperature. Fitting $\ell_{\phi}^{-2} \propto T^{p}$ yields $p$ between 2 and 3 in all magnetic phases, consistent with electron-phonon scattering.

Power-law exponents extracted from data ranging over a single decade in temperature likely have significant quantitative uncertainties. This challenge is exacerbated by the low-temperature saturation in $\ell_{\phi}$ apparent in Figure~\ref{fig5}(d). Similar behavior is also visible in Figure~\ref{fig5}(e), which shows $R_{xx}$ as a function of temperature at 1.5 T (AFM phase), 5.5 T (cAFM phase), and 11.5 T (FM phase). The saturation of transport behavior below about 100~mK was likely caused by current-driven heating resulting in electron temperature decoupling from the thermometer-measured lattice temperature. We explore this further in Section~S6~\cite{sup}. 

In each magnetic phase, Figure~\ref{fig5}(e) reveals $R_{xx} \propto -\log(T)$ for temperatures above this saturation. Common causes of a low-temperature logarithmic increase in $R_{xx}$ as temperature decreases include weak localization~\cite{altshuler1980} or Kondo~\cite{kondo1964} effects. Each of these mechanisms is suppressed by magnetic field: weak localization at a scale $B\sim B_c$ and Kondo at a scale $B\sim k_B T_K/(g \mu_B)$ (here $k_B$ is the Boltzmann constant, $g$ is the $g$-factor, and $\mu_B$ is the Bohr magneton). All the data of Figure~\ref{fig5}(e) were acquired at $B\gg B_c$, excluding weak localization as a possible explanation for the temperature scaling. As for Kondo, we see no significant field dependence for temperature scaling of resistivity out to our maximum field of 12~T. To avoid suppression of Kondo physics across this field range, the Kondo temperature $T_K$ would need to be at least several times larger than $12~\mathrm{T} * g\mu_B / k_B \approx 16$~K. But scattering from Kondo impurities shows logarithmic scaling only for $0.1 T_K \lessapprox T \lessapprox 10 T_K$, so explaining our resistivity data with such a large Kondo temperature is not internally consistent with our observation of logarithmic scaling down to a fraction of a Kelvin. Instead, we attribute this scaling to another well-established phenomenon: the low-temperature suppression of the density of states near the Fermi level caused by electron-electron interactions~\cite{aa1985,butko2000}.


Having established the basic phenomenology of UCFs in our MBT device, we next explore the possible relationship between features of our data and domains in the magnetization of MBT. Magnetic domains in the AFM and cAFM phases of exfoliated flakes of MBT have been studied by magnetic force microscopy~\cite{sass2020,sass2020b}. Antiferromagnetic domains result from intralayer regions of opposite magnetization; canted antiferromagnetic domains result from intralayer regions of opposite canting~\cite{sass2020}. In contrast, in the FM phase domains are eliminated by the high external magnetic fields and MBT is found to be uniformly magnetized~\cite{sass2020}. Although the presence of domains is well-established through these magnetic force microscopy measurements, their impact on transport measurements has yet to be studied. 

We suggest that the field-sweep-direction-dependent changes to the UCF patterns in the AFM and cAFM phases (Figure~\ref{fig3}) indicate reversible motion of magnetic domain walls as a function of applied field. Such an interaction between magnetic domain reconfiguration and UCF patterns should be expected. Variations in both elastic and inelastic charge carrier scattering upon changing the magnetic domain structure are well-known theoretically~\cite{tatara1997,lyanda1998,brataas1999} and experimentally~\cite{hong1995,gregg1996,hong1996,ruediger1998,aumentado2000}. Since UCFs are sensitive to changes in the scattering landscape seen by charge carriers, the magnetofingerprint should therefore be sensitive to changes in the magnetic domain structure. Consistent with this interpretation, micron-scale field-driven domain wall motion has been observed via magnetic force microscopy~\cite{sass2020}, although correlations across repeated field sweeps or as sweep direction is reversed have not previously been examined.

Further signatures of interaction between magnetic domain walls and UCF patterns can be seen in changes to the UCF pattern upon crossing magnetic phase boundaries, as shown in Figure~\ref{fig4}. A spin-flop transition occurs at the AFM/cAFM phase boundary~\cite{sass2020b,bac2022,liu2022}. The number and location of magnetic domain walls need not be preserved through this phase transition, and different domain patterns on either side of this phase transition have been observed in magnetic force microscopy~\cite{sass2020}. We suggest the reduction in $\R$ upon crossing the AFM/cAFM magnetic phase boundary indicates the  formation of new domains, and therefore a new scattering landscape. In contrast, crossing the cAFM/FM phase boundary does not appear to substantially perturb the UCF pattern, consistent with reports that crossing this boundary brings not a sudden rearrangement of magnetization but rather a gradual reduction of the in-plane canting angle~\cite{bac2022}.

Taken together, these observations suggest measurements of UCF serve as a transport-based probe of magnetic domain wall reconfiguration in MBT thin films. Though our experiment cannot directly measure correlation between domain wall positions and UCF magnetofingerprints, we are aware of no alternative mechanism by which the scattering landscape of MBT could exhibit the observed field-sweep-direction-dependence in the AFM and cAFM phases, but not the FM phase. 
Future simultaneous scanning probe and transport measurements could more rigorously test the relationship between domain wall motion and UCF measurements. Sensitivity to magnetic ordering on a 100 nm length scale while in a high magnetic field is within the capabilities of magnetic force microscopy and (at least for the AFM phase) of nanoSQUID~\cite{finkler2010}. 

In this work, we present measurements of universal conductance fluctuations in a thin film of the magnetic topological material \ce{MnBi2Te4}. Analysis of the UCFs in three magnetic phases of MBT reveal information on phase coherence lengths and coherence-breaking scattering mechanisms. We observe strong indications that magnetic domain wall motion and domain rearrangement are expressed in UCF measurements. Our work establishes UCFs as a transport-based probe of magnetization heterogeneities and spin-flop transitions in a topological magnet, a first step toward the critical experimental need for practical probes relating topological and magnetic order in disordered materials.


The authors thank D. Natelson, B. Shklovskii, and L. K. Rodenbach for useful discussions. M. P. A., E. M., I. T. R., M. A. K., and D. G.-G. were supported by the U.S. Department of Energy, Office of Science, Basic Energy Sciences, Materials Sciences and Engineering Division, under Contract DE-AC02-76SF00515. M. P. A. additionally acknowledges support from Gordon and Betty Moore Foundation through Grant No. GBMF9460 and Air Force Office of Scientific Research (AFOSR) Multidisciplinary Research Program of the University Research Initiative (MURI) under grant number FA9550-21-1-0429 during the later stages of this work. L.T., P.Z., and K.L.W. acknowledge support from the National Science Foundation (NSF) under grant numbers 2125924 and 1936383 and the Army Research Office Multidisciplinary University Research Initiative (MURI) under grant number W911NF-19-S-0008. Infrastructure and cryostat support were funded in part by the Gordon and Betty Moore Foundation through Grant No. GBMF3429. We thank NF Corporation for providing low-noise, high-input-impedance voltage preamplifiers.  We acknowledge measurement assistance from colleagues at the National Institute of Standards and Technology. Part of this work was performed at the nano@Stanford labs, supported by the National Science Foundation under award ECCS-2026822. 

\setlength\bibitemsep{0pt}
\printbibliography

\titleformat{\section}
   {\normalfont\Large\bfseries}{\thesection}{1em}{}
\titleformat{\subsection}
   {\normalfont\large\bfseries}{\thesubsection}{1em}{}

\renewcommand{\thefigure}{S\arabic{figure}}
\renewcommand{\theequation}{S\arabic{equation}}
\renewcommand{\thesection}{S\arabic{section}}

\newpage
\onecolumn

  \begin{flushleft}
        \Large\textbf{Supplemental information for: Universal conductance fluctuations in a \ce{MnBi2Te4} thin film}
    \end{flushleft}

    \begin{flushleft}
        {Molly P. Andersen$^{1,2}$, Evgeny Mikheev$^{3,2,4}$, Ilan T. Rosen$^{5,2,6}$, Lixuan Tai$^7$, Peng Zhang$^7$, Kang L. Wang$^7$, Marc A. Kastner$^{3,2,8}$, and David Goldhaber-Gordon$^{3,2,a}$}
    \end{flushleft}
    
    \begin{flushleft}
        \footnotesize{$^1$\textit{Department of Materials Science and Engineering, Stanford University, Stanford, California 94305, USA}}\\
        \footnotesize{$^2$\textit{Stanford Institute for Materials and Energy Sciences, SLAC National Accelerator Laboratory, 2575 Sand Hill Road, Menlo \taba Park, California 94025, USA}}\\
        \footnotesize{$^3$\textit{Department of Physics, Stanford University, Stanford, California 94305, USA}}\\
        \footnotesize{$^4$\textit{Department of Physics, University of Cincinnati, Cincinnati, Ohio 45221, USA}}\\
        \footnotesize{$^5$\textit{Department of Applied Physics, Stanford University, Stanford, California 94305, USA}}\\
        \footnotesize{$^6$\textit{Research Laboratory of Electronics, Massachusetts Institute of Technology}}\\
        \footnotesize{$^7$\textit{Department of Electrical and Computer Engineering, Department of Physics and Astronomy, University of California, Los Angeles, California 90095, USA}}\\    
        \footnotesize{$^8$\textit{Department of Physics, Massachusetts Institute of Technology, Cambridge, Massachusetts 02139, USA}}\\    
        \footnotesize{$^a$To whom correspondence should be addressed; E-mail: \texttt{goldhaber-gordon@stanford.edu}} \newline \newline \newline \newline
    \end{flushleft}

\renewcommand{\thetable}{S\arabic{table}}
\renewcommand{\thefigure}{S\arabic{figure}}
\renewcommand{\thesection}{S\arabic{section}}
\renewcommand{\thesubsection}{\roman{subsection}}
\renewcommand{\theequation}{S\arabic{equation}}

\setcounter{secnumdepth}{3}
\setcounter{equation}{0}
\setcounter{figure}{0}
\setcounter{section}{0}
\setcounter{tocdepth}{1}

\tableofcontents

\clearpage

\section{Details of experimental methodology}

The \ce{MnBi2Te4} film used in this work~[S1] was nominally 6 septuple layers (SLs) thick and was grown in a Perkin-Elmer ultra-high vacuum molecular beam epitaxy (MBE) system. Throughout the process, growth was monitored by \textit{in situ} reflection high-energy electron diffraction (RHEED), with RHEED images captured by a KSA400 system from K-space Associates, Inc. The growth substrate was epi-ready semi-insulating GaAs (111)B; after loading the substrate into the MBE chamber, substrates were pre-annealed at 630~\degree C in a Te-rich environment. The substrate was kept at 200~\degree C during growth. To grow the film, high-purity Mn, Bi and Te were evaporated simultaneously from standard Knudsen cells. The film was annealed in a Te-rich environment for 2 minutes at 290~\degree C after deposition to improve crystallinity.

\begin{figure*}[h]
\centering
	\includegraphics[width=17cm]{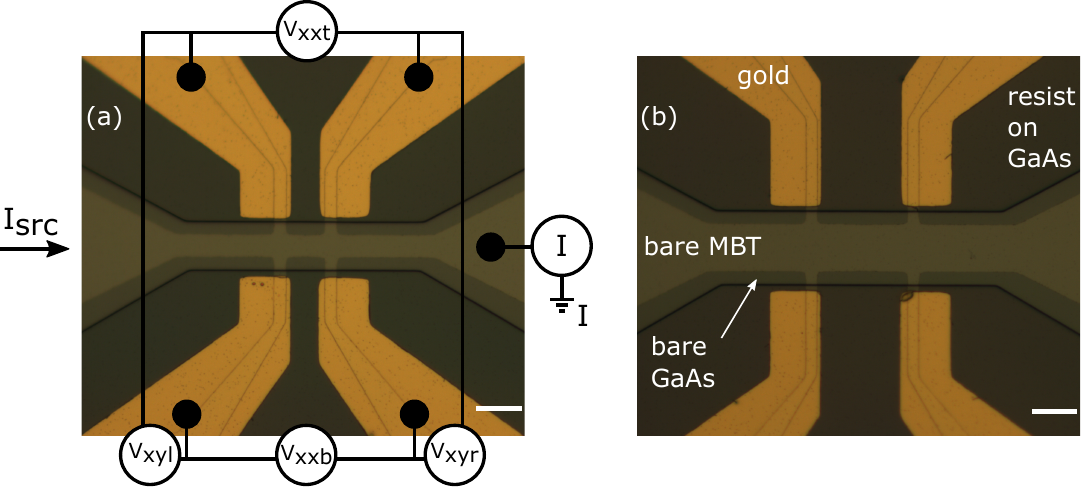}
	\caption{Optical micrographs of (a) Device A, with two 10~$\mu$m squares between longitudinal voltage taps, and (b) Device B, with two 20~$mu$m squares between longitudinal voltage taps. Electrical measurement setup is indicated in (a): Current $I_{src}$ is sourced in from the left, measured ($I$), and drained to ground out the right contact. All combinations of longitudinal ($V_{xxt}$, top longitudinal pair; $V_{xxb}$, bottom longitudinal pair) and Hall ($V_{xyl}$, left Hall pair; $V_{xyr}$, right Hall pair) voltages were measured as indicated. Device B was measured in the same manner. Components of the optical micrographs are described in (b); both images were taken after the device mesa was defined (bare MBT region), contacts were added (gold region), a global ALD dielectric was deposited, and the top gate electrodes were patterned with photolithography. Photoresist from the top gate electrode lithography step is still in place in both images (resist on GaAs). Scale bars in both images: 20~$\mu$m.}
	\label{sfig-devices}
\end{figure*}

Two Hall bar devices were fabricated on a single chip of this film. To define device mesas, the chip was first cleaned in acetone and isopropanol. After patterning mesa geometries with photolithography, the mesas were etch-defined with 35 s Ar ion milling with a 400 V accelerating voltage. The chip was cleaned with 30 s sonication in acetone and an isopropanol rinse to remove resist and sidewalls from ion milling. Current and voltage contacts were then defined with photolithography, and 5/85 nm Ti/Au was deposited with electron beam (e-beam) evaporation after a 10 s \textit{in situ} Ar pre-etch. Contacts were lifted off with a 5 minute soak in acetone and 10 s sonication followed by a solvent rinse. An alumina electrostatic top gate was then added. First, a 1 nm seed layer of aluminum was globally deposited by e-beam evaporation and allowed to oxidize in air. Next, 400 cycles of atomic layer deposition (ALD) was used to deposit alumina with tetramethylammonium and water precursors at 60~\degree C, where the reduced temperature was chosen to avoid thermal degradation of the sample. The chip was next rinsed with acetone and isopropanol. A photolithographically-masked 150 s wet etch with Microposit MF-CD-26 developer was used to remove alumina over the contact pads. The chip was again rinsed in acetone and isopropanol to remove the photoresist. Top gate electrodes were patterned with photolithography; at this point optical micrographs of the devices were taken (Figure~\ref{sfig-devices}). E-beam evaporation was then used to deposit 5/95 nm Ti/Au after a 10 s \textit{in situ} Ar pre-etch. Metal liftoff was performed with sonication in acetone and a solvent rinse. 

All photolithography steps were performed as follows: The sample was spin-coated at 5,500 rpm with a hexamethyldisilazane adhesion layer and SPR 3612 photoresist. The resist was then baked at 80~\degree C for 5 minutes; this lower-than-typical bake temperature was chosen to avoid thermal decomposition of the film. Patterns were exposed on a MicroWriter Direct Write tool with a 385 nm wavelength and a 115 mJ/cm$^2$ dose. Exposed regions of the resist were developed with 35 s/20 s/20 s rinses in Microposit Developer CD-30/deionized (DI) water/DI water. 

Electronic transport measurements were performed in a Leiden Cryogenics dilution refrigerator with a base temperature of about 35 mK. Measurement lines were filtered with low-pass RF filters and discrete RC filters at the mixing chamber stage to ensure charge carrier thermalization. A 5-8 Hz ac current was sourced separately into each device by a Stanford Research 830 lock-in amplifier (SR830) applying a 2 or 5 V signal across a 1 G$\Omega$ resistor to produce a 2 or 5 nA bias current for main text data. Voltage measurements were acquired by additional SR830s, as well as Stanford Research 860 lock-in amplifiers, after $10^2$ V/V amplification by either separate NF LI-75A voltage preamplifiers or a single NF multi-channel preamplifier. Currents were measured on SR830s after amplification by an Ithaco 1211 current preamplifier. External out-of-plane magnetic fields were applied using a 14~T superconducting magnet. Measurements at elevated temperatures were taken by heating the sample with a heater on the mixing chamber stage of the dilution refrigerator. Measurements were taken between two separate cool-downs. Table~\ref{table1} shows the magnitude of current excitations and step sizes in applied field for main-text Figures~2--5. Larger currents were used during measurements acquired in the first cool-down, before the impact of the current excitation magnitude had been realized. Larger step sizes in applied field were used for main-text Figure~2 to limit measurement time, but smaller step sizes were used for all other measurements. 
\begin{center}
\label{table1}
\begin{tabular}{| c | c | c  |}
\hline
Main-text figure & Current excitation (nA) & Applied field step size (mT) \\ 
\hline
2 & 5 & 5 \\  
3 & 2 & 2 \\
4 & 5 & 2 \\
5 & 2 & 2\\    
\hline
\end{tabular}
\end{center}

\clearpage

\section{Additional linecuts of atomic force microscopy images}

\begin{figure*}[h]
\centering
	\includegraphics[width=17cm*1/3]{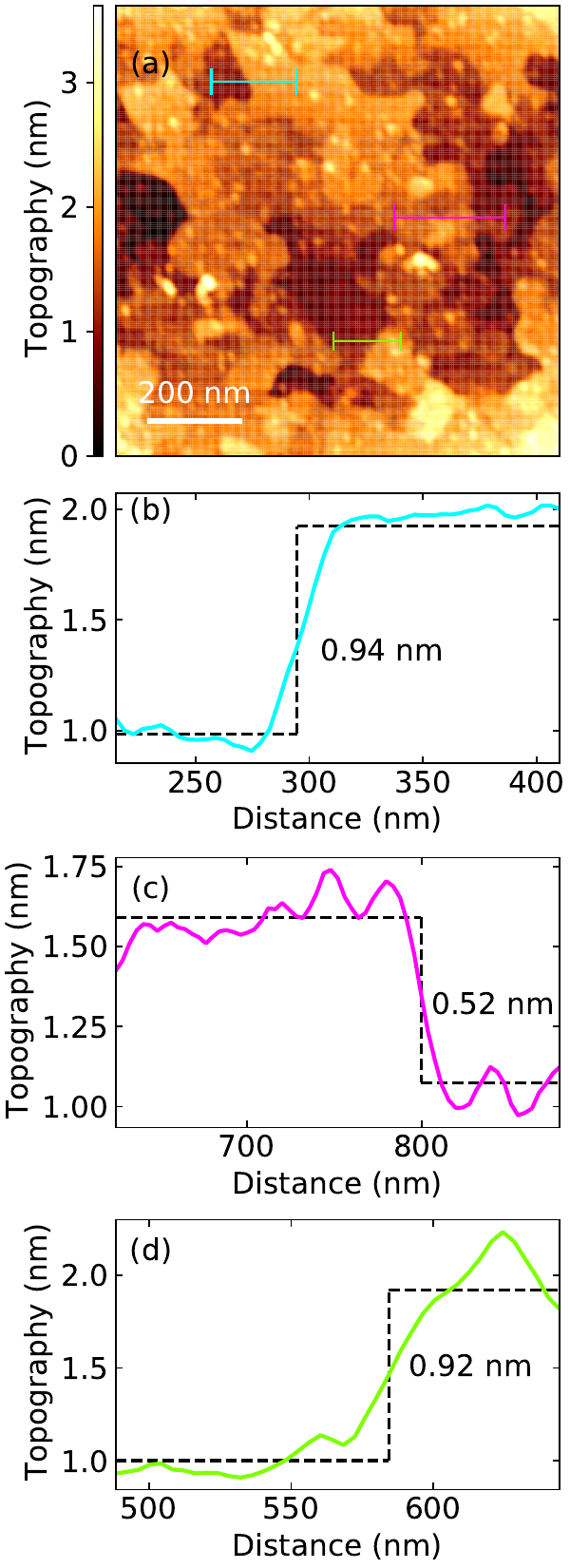}
	\caption{(a) Reproduction of Figure~1(b), showing a topographic map of the MBT thin film acquired with atomic force microscopy. (b-d) Additional linecuts with extracted step heights. The location of each linecut in (a) is are indicated by (b) cyan, (c) fuchsia, and (d) chartreuse lines. Data were averaged across several parallel traces in the transverse direction, indicated by the extent of the crosspieces appended to the tails of the lines in (a).}
	\label{sfig-afm}
\end{figure*}

Figure~\ref{sfig-afm} shows three additional linecuts across step edge features in the topographic map of the MBT thin film. Two of these steps were about 0.9~nm, consistent with \ce{Bi2Te3}. One step is 0.5~nm, possibly resulting from \ce{MnTe_x}.

\clearpage

\section{Density, mobility, and thermal diffusion length}

\begin{figure*}[h]
\centering
	\includegraphics[width=17cm*2/3]{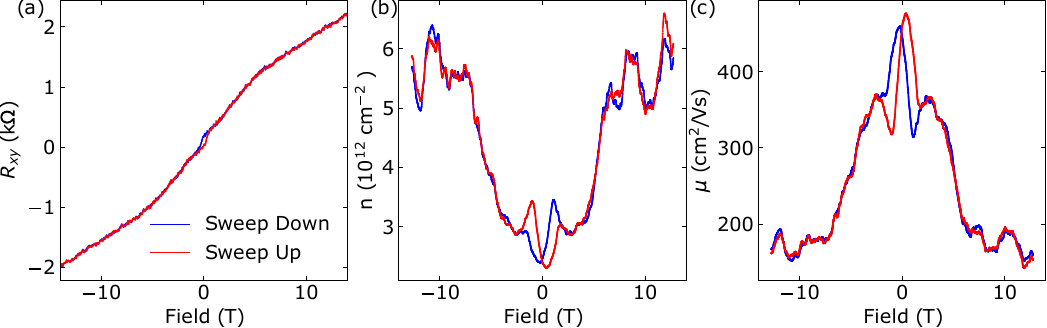}
	\caption{(a) Hall measurements acquired simultaneously with the data of Figure~\ref{fig2}. (b) Field-dependent density $n$ extracted from the Hall measurements of (a) over a 2.5 T sliding window. (c) Mobility $\mu$ derived from the longitudinal and Hall measurements of Figure~\ref{fig2} and (a). For all data, red (blue) points were acquired with field sweeping up (down)}
	\label{sfig-hall}
\end{figure*}

Hall resistance measured as a function of external magnetic field on the left voltage contact pair on Device A is shown in Figure~\ref{sfig-hall}(a). Electron density $n$, extracted from the slope of the Hall resistance over a 2.5 T sliding window, is shown in Figure~\ref{sfig-hall}(b). At lower external magnetic fields, a higher Hall slope (and correspondingly lower calculated charge carrier density) is observed. These changes in Hall signal are likely related to MBT's magnetic phase transitions, as discussed in the context of longitudinal resistance measurements in the main text. We therefore opt to use densities extracted near zero external magnetic field ($2.5 * 10^{12}$ cm$-2$) for our calculations. Mobility $\mu$, calculated via $\sigma_{xx} = ne\mu$ for longitudinal conductivity $\sigma_{xx}$, is shown in Figure~\ref{sfig-hall}(c). 

Using the calculated mobility, $l_{T}=\sqrt{hD/k_{B}T}$ can be approximately determined. Diffusivity $D$ can be calculated as $D=v_F\ell_{el}/4=v_F^2 \mu m^{*}/4e$, where $l_{el}$ is the elastic scattering length. With a minimum mobility $\mu \approx 200$~cm$^2$/Vs (Figure~\ref{sfig-hall}(b)), electron charge $e$, and estimates for effective mass $m^*=0.25$~electron mass and surface state Fermi velocity $v_F=3*10^5$~m/s taken from theory and other experiment~[S2,S3], $l_T \approx 2$~$\mu$m at 30 mK for our MBT thin film. We note the band structure details used to calculate diffusivity are only approximate, but our estimation of $l_T$ is likely correct to within an order of magnitude.

\clearpage

\section{Correlation in magnetic field}

\begin{figure*}[h]
\centering
	\includegraphics[width=17cm*2/3]{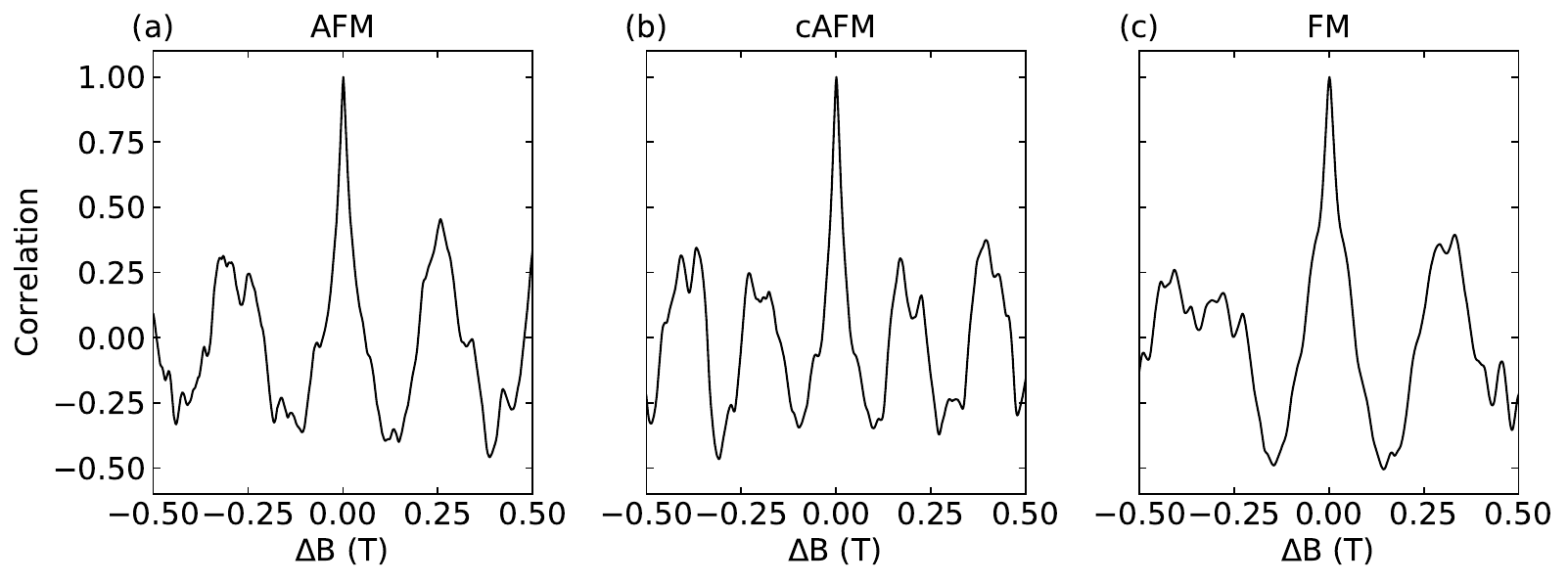}
	\caption{Correlation $C(\Delta B)$ calculated from conductance fluctuations measured across a 1~T range with 2~mT step size and a 2~nA bias current in the (a) AFM, (b) cAFM, and (c) FM phases. Conductance fluctuation data from which these plots were calculated are shown in ``Down 1'' sweeps of Figure~\ref{fig2}.}
	\label{sfig-correlation}
\end{figure*}

As described in the main text, an analysis of the field scale over which correlations persist in UCFs can provide insight into $l_{\phi}$. This method provides a sensible estimate of $l_{\phi}$ when applied to large-field-domain data in Figure~\ref{fig2}. However, most data acquired in this work was taken over a 1~T field range. Calculated $C(\Delta B)$ plots for these 1~T-range datasets reveal a central coherence peak surrounded by reasonably high-intensity quasi-periodic oscillations instead of minimal noise about $\Delta B=0$~T as is typical. These oscillations appear to affect the central coherence peak and therefore prevent the analysis of $B_c$ described in the main text. A few examples are shown in Figure~\ref{sfig-correlation}. 

One possible explanation for these qualitative features is an actual periodic signal is obscured below the apparent random conductance fluctuations. Such a periodic signal could arise from the Aharanov-Bohm effect~[S4], caused by quantum interference of charge carriers traversing rings with flux threaded through the ring. If the Aharanov-Bohm effect were present in our data, different analysis techniques than are presented in the main text would be required.

\subsection{Fast Fourier transform analysis}

To show the low frequency quasi-periodic oscillations in the correlation functions did not arise from currents forming Aharanov-Bohm loops, we compare the fast Fourier transforms (FFTs) of UCFs over subsets of the field range plotted in Figures~\ref{fig2} and~\ref{sfig-hall}. The full dataset was acquired in 5 nA steps while sweeping magnetic field from 14~T$\rightarrow$-14~T$\rightarrow$14~T at base temperature. Here, we limit our analysis to the $|10-14|$~T ranges for both field sweep directions, where it can be assumed there is no impact from magnetic domain wall motion. Different quasi-periodic oscillations in different field ranges indicates that these oscillations are not associated with specific Aharanov-Bohm loops but rather an artifact of a limited field range.

\begin{figure*}[h]
\centering
	\includegraphics[width=17cm*5/6]{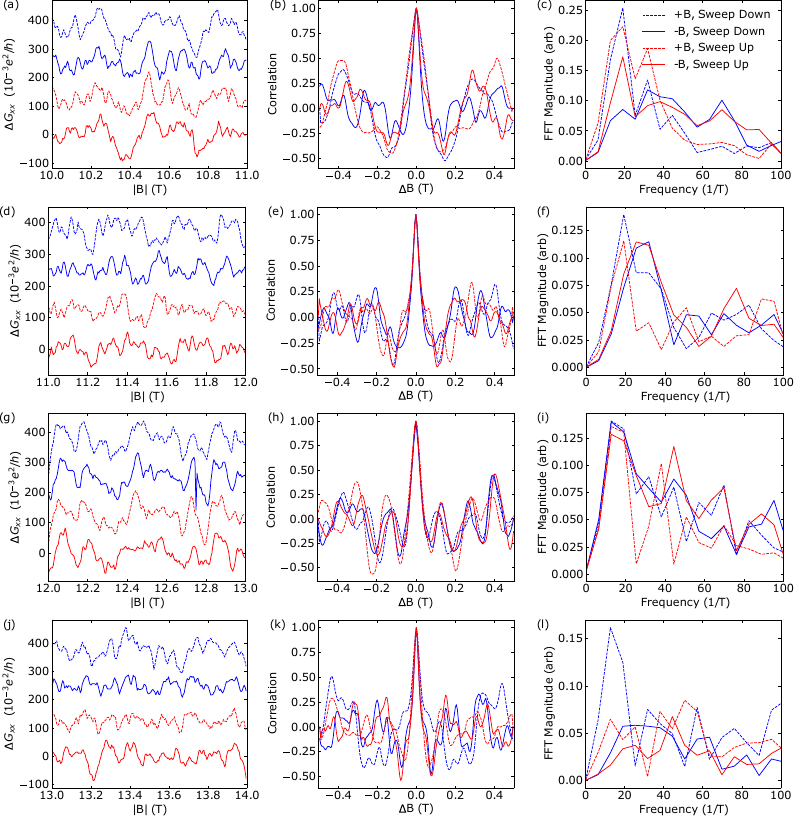}
	\caption{Conductance fluctuations, correlation function, and fast Fourier transform magnitude for data taken over 1~T field ranges between 10~T and 14~T. (a-c) 10-11~T; (d-f) 11-12~T; (g-i) 12-13~T; (j-l) 13-14~T. Blue (red) traces were acquired with external field sweeping down (up). Dashed (solid) lines were acquired with a positive (negative) out-of-plane external magnetic field.}
	\label{sfig-fft-1T}
\end{figure*}

\begin{figure*}[h]
\centering
	\includegraphics[width=17cm]{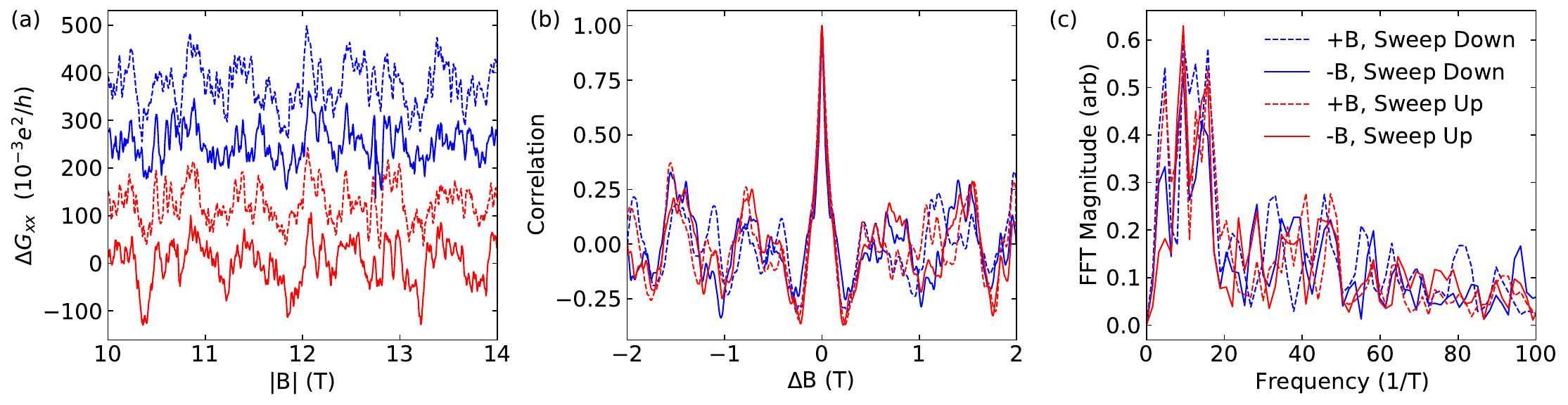}
	\caption{Conductance fluctuations, correlation function, and fast Fourier transform magnitude for data taken over a 4~T field range between 10~T and 14~T. Blue (red) traces were acquired with external field sweeping down (up). Dashed (solid) lines were acquired with a positive (negative) out-of-plane external magnetic field.}
	\label{sfig-fft-4T}
\end{figure*}

Figures~\ref{sfig-fft-1T} and~\ref{sfig-fft-4T} plot conductance fluctuations, calculated correlation, and FFTs of the conductance fluctuations for 1~T and 4~T field ranges. Both field sweep directions and field orientations are included in each plot, resulting in four unique traces per plot. Conductance fluctuations and correlation functions are calculated from measurements as discussed in the main text. The FFTs are calculated from $\Delta G_{xx}$ traces as follows: (1) windowing the data with a Hamming window; (2) assuming only real inputs, calculating a discrete one-dimensional FFT with Numpy's built-in routine; (3) calculating the frequency-dependent magnitude of the FFT by combining the real and imaginary parts of the FFT; (4) normalizing the magnitude by a constant to bring the maximum FFT amplitude close to unity.

For the smallest field domain (1~T, Figure~\ref{sfig-fft-1T}), some low-frequency quasi-periodic oscillations are clear in the correlation plots (for example, black-boxed regions). Importantly, different field ranges result in apparently distinct oscillations. If Aharanov-Bohm loops were present one expects the same underlying oscillations to persist at all field ranges. Additionally, as the field domain is increased, the magnitude of any specific oscillations are suppressed and the correlation function appears random outside of the central coherence peak (Figure~\ref{sfig-fft-4T}). If oscillations persisted throughout the whole field range, their magnitude would not be suppressed by extending the field domain. This again suggests transport around Aharanov-Bohm loops are not the source of the quasi-periodic oscillations in some correlation field plots.

However, although visual inspection of the correlation analyses appears to rule out Aharanov-Bohm effects, noise data requires more careful examination. Here, we turn to the FFTs of the $\Delta G_{xx}$ data. Across the field range and domains studied here no consistent low-frequency structure is apparent. Substantial variation is clear even between data related by Onsager relations (i.e. traces within the same plot). This supports the conclusion that the conductance fluctuations we observe are caused by universal conductance fluctuations and not by currents circulating around a network of loops. 

\clearpage

\section{Repeatability over long timescales}

\begin{figure*}[ht]
\centering
	\includegraphics[width=17cm*5/6]{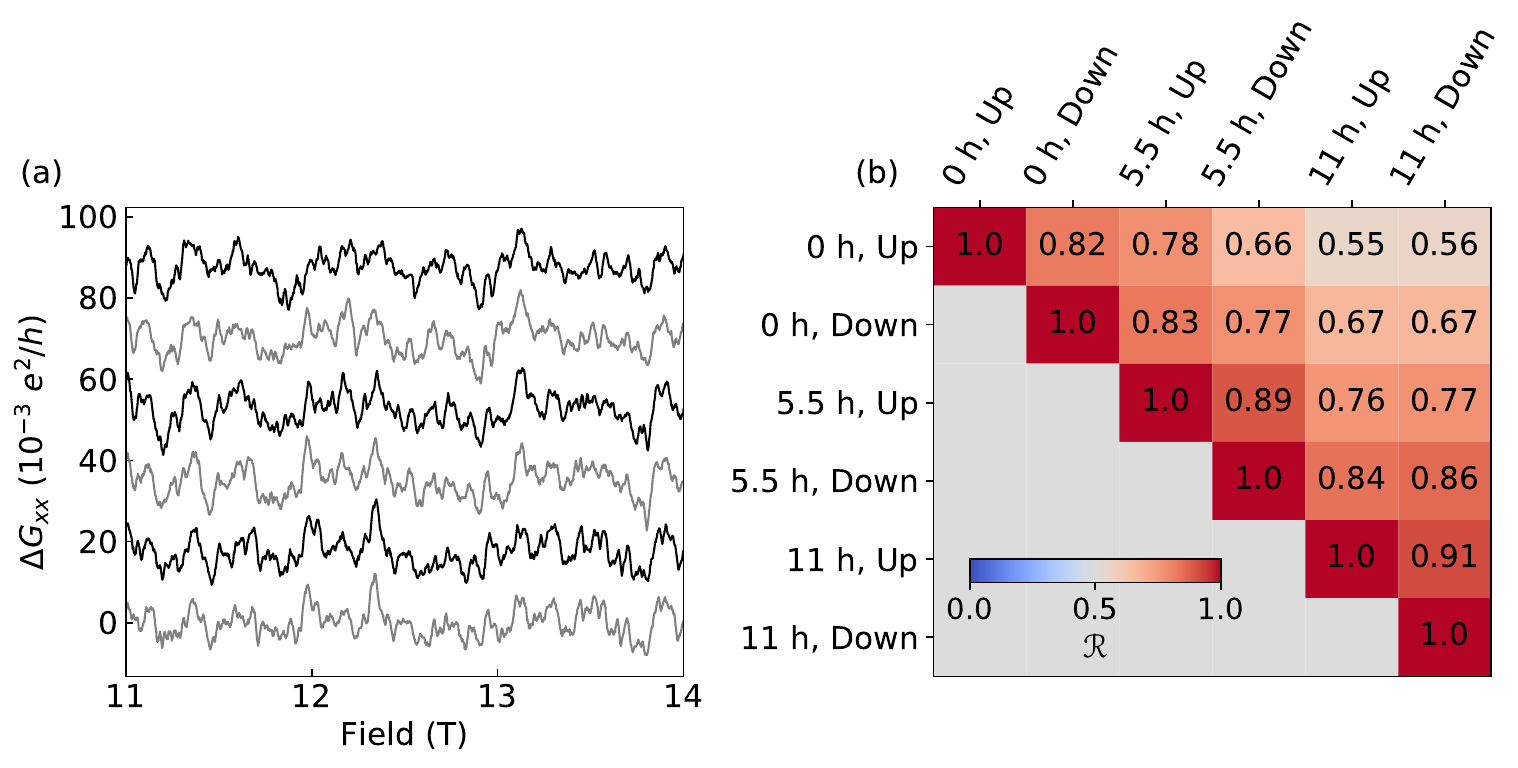}
	\caption{Repeatability of conductance fluctuations in the FM phase across a  14~hour time window. (a) Conductance fluctuations acquired between 11 and 14~T with a 2~mT step size and a 5~nA bias current. Traces shown in black (grey) were acquired while sweeping field up (down). Traces were acquired sequentially starting from the top and are offset for clarity. (b) Repeatability $\R$ for the data shown in (a).}
	\label{sfig-time}
\end{figure*}

Figure~\ref{sfig-time} demonstrates the repeatability of the UCFs as a function of external magnetic field. To acquire the conductance fluctuation data shown in (a), the magnetic field was swept 11~T$\rightarrow$14~T$\rightarrow$11~T. Each 3~T field sweep took about 100~minutes. After a 2~hour delay at 11~T, the pair of field sweeps were repeated. After a second 2~hour delay, the pair of field sweeps were repeated again. All data was acquired with a 5~nA current bias. The correlation between pairs of field sweeps is presented in Figure~\ref{sfig-time}(b). For successive runs, $\R > 0.8$. For runs separated by about 11~hours, $\R>0.5$. 

We note that the highest values of $\R$ are smaller in the data shown in Figure~\ref{fig4} than that shown in Figure~\ref{fig3}. We attribute this reduction to the longer data collection time (limited by the magnet's ramp rate) of measurements shown in Figure~\ref{fig4}: the data shown each sub-panel of Figure~\ref{fig3} were collected in about 3.5 hours, while the complete measurement shown in Figure~\ref{fig4}(a) took about 29 hours. We believe rearrangement of disorder on the timescale of the latter measurement reduces the measured correlation. Additionally, measurement parameters like current bias, domain in field, and step size in field can affect $\R$.

\clearpage
\section{Varied ac current bias}

\begin{figure*}[h]
\centering
	\includegraphics[width=17.2cm*2/3]{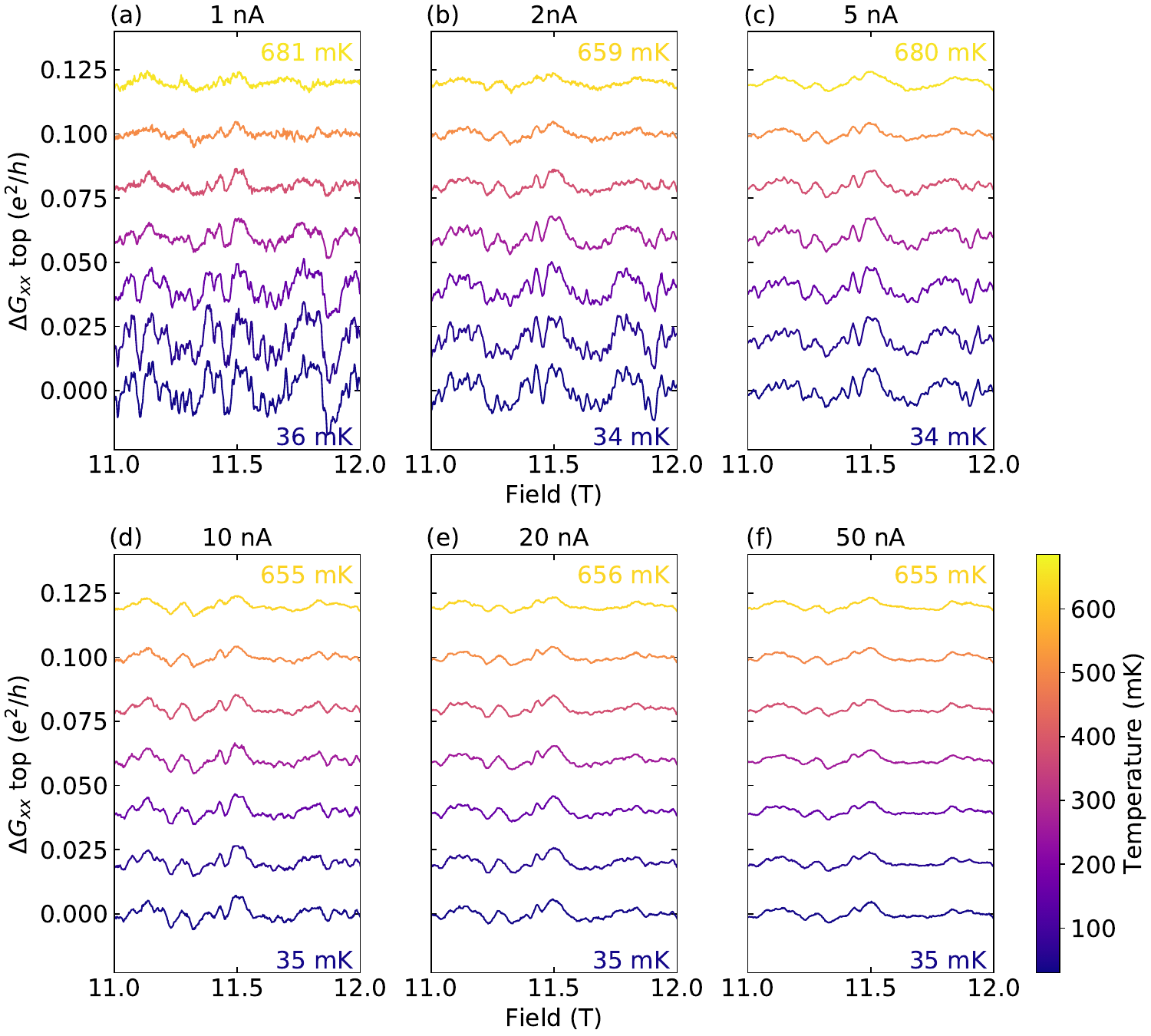}
	\caption{Artificially offset traces of $\Delta G_{xx}$ from the top contact pair of Device A, acquired as field is swept between 11 and 12 T at varied temperatures in the same manner as the data presented in Figure~\ref{fig4} (2 nA data shown here is reproduced in Figure~\ref{fig4}). Different subplots were acquired with differing ac current biases: (a) 1 nA; (b) 2 nA; (c) 5 nA; (d) 10 nA; (e) 20 nA; (f) 50 nA. As current bias is increased, the magnitude of UCF at low temperatures decreases.}
	\label{sfig-current-temp}
\end{figure*}

A low-temperature saturation in both $R_{xx}$ and $\Delta G_{xx}$ is observed in the main-text data (see Figure~\ref{fig4}(d,e)). Several possible explanations exist for this behavior, including scattering lengths exceeding device dimensions (unlikely, given that the device is 100 times larger than the largest $l_{\phi}$ seen here) or Kondo physics (previously excluded due to the lack of magnetic field dependence). Instead, we observe the saturation to be an artifact of ac current-driven Joule heating of the sample. Figure~\ref{sfig-current-temp} shows similar measurements to those in main text Figure~\ref{fig4}, taken within the FM phase with ac current biases ranging from 1-50 nA. As current bias is increased, the magnitude of lowest-temperature $\Delta G_\mathrm{rms}$ clearly decreases, indicating heating of the sample and therefore a smaller phase coherence length. By 50 nA, increasing temperature from 35-680 mK causes only minimal changes in the UCF pattern, which suggests the effective sample temperature is at least several hundred milliKelvin.

\begin{figure*}[h]
\centering
	\includegraphics[width=17cm]{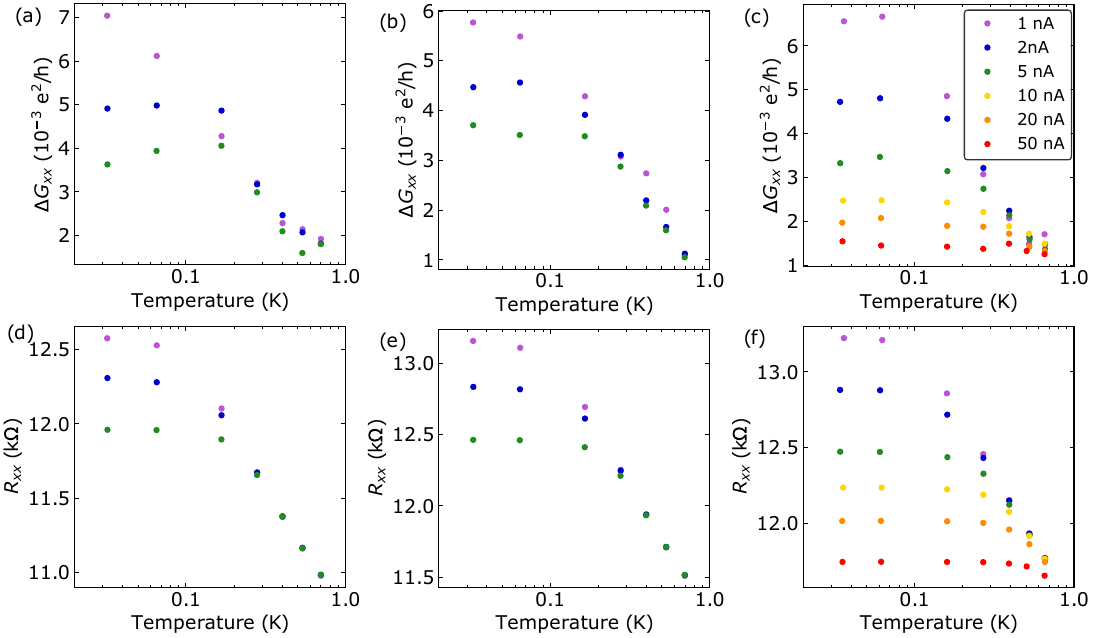}
	\caption{Temperature dependence of (a-c) $\Delta G_\mathrm{rms}$ and (d-f) $R_{xx}$ in the (a,d) AFM (b,e) cAFM and (c,f) FM magnetic phases with varied ac current bias. All data was acquired with the top contact pair of Device A.}
	\label{sfig-current-summary}
\end{figure*}

Figure~\ref{sfig-current-summary}(c,f) shows analogies of the main text Figure~\ref{fig4}(d,e) for the FM phase and varied current bias. Low-temperature saturation in both $R_{xx}$ and $\Delta G_\mathrm{rms}$ clearly increases with the ac current bias, a clear indication of current-driven Joule heating. Similar measurements at a few low current biases demonstrate the same behavior in both the AFM and cAFM phases; a summary of data in the AFM and cAFM phases is shown in Figure~\ref{sfig-current-summary}(a,b,d,e). 

We note that our main text data was acquired with either a 2 or 5~nA current bias. At these currents, this current-driven heating effect is apparent at lowest temperatures, resulting in suppressed measurements of $\Delta G_\mathrm{rms}$. It is therefore likely we underestimate $\ell_{\phi}$ in our lowest-temperature calculations. A similar measurement suite with ac currents below 1~nA is necessary to more accurately determine $\ell_{\phi}$. 

Additionally, this current-driven heating decouples the lattice temperature from the electron temperature. The thermometer on the mixing chamber stage of the probe therefore does not accurately reflect the temperature of the charge carriers. Effective electron temperature can be estimated from Figure~\ref{sfig-current-summary} as the temperature at which $\Delta G_{xx}$ and $R_{xx}$ are no longer saturated and begin changing with increasing temperature. For 2 and 5~nA bias currents, this appears to be 80-120~mK. We note that the elevated electron temperature relative to lattice temperature does not affect our argument that $\ell_T >> \ell_{\phi}$. Since $\ell_T \propto T^{-1/2}$, we expect at most a $\sqrt{3}$ reduction in $\ell_T$ compared to what we calculated above. In this case, $\ell_T \approx 1$~$\mu$m, which still exceeds our highest estimates of $\ell_{\phi}$.

\clearpage

\section{Electrostatic gating}

\begin{figure*}[h]
\centering
	\includegraphics[width=17.cm*1/2]{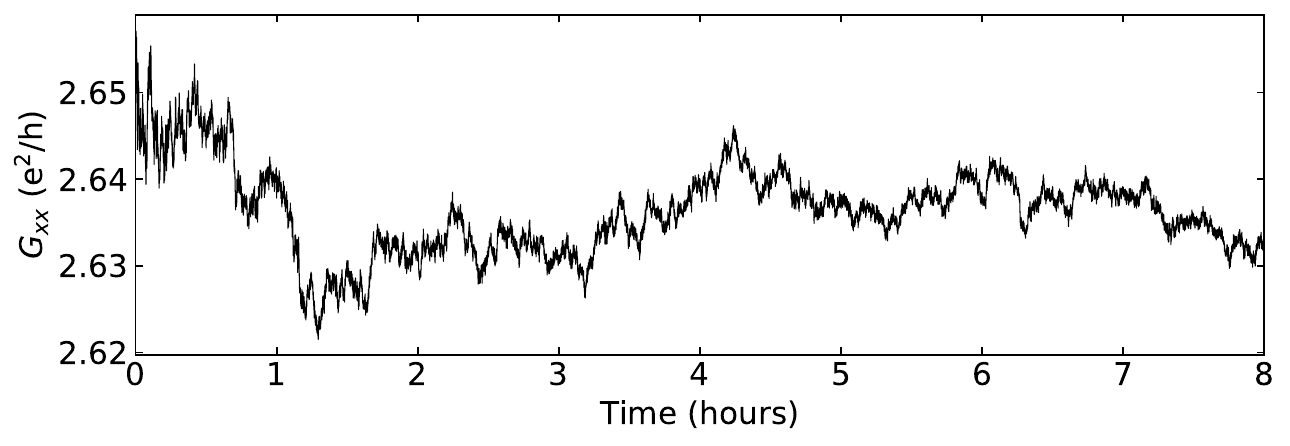}
	\caption{Longitudinal conductance of the 10 $\mu$m Hall bar as a function of time after last gate voltage excursion.}
	\label{sfig-gate}
\end{figure*}

As mentioned above, Hall bars presented here featured electrostatic top gates to allow experimental control over charge carrier density. However, long timescales for equilibration of charge traps after changes to the applied gate voltage restrict this work to data acquired exclusively with the top gate grounded. Figure~\ref{sfig-gate} shows measured longitudinal conductance $G_{xx}$ over time after the top gate voltage is changed from -4~V $\rightarrow$ 4~V. Persistent random fluctuations in $G_{xx}$ on the order of 0.01~$e^2/h$ are obvious. We attribute these fluctuations to slow (de-)population of charge traps within the gate dielectric (likely, since the dielectric was deposited at low temperatures (see above) and therefore contains residual carbon contamination and other defects) or elsewhere in the system. Since the electrostatic landscape contributes to the elastic scattering landscape charge carriers see, these charge trap (de-)population events affect the device conductance. 

\begin{figure*}
\centering
	\includegraphics[width=5cm]{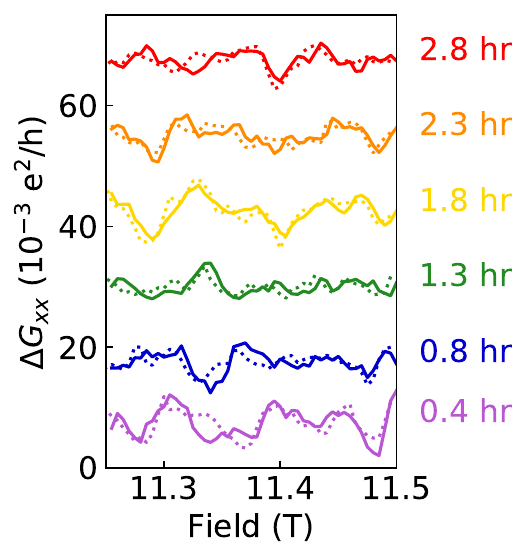}
	\caption{Conductance fluctuations as a function of external magnetic field with varied start times after the last gate voltage excursion. Solid (dotted) lines correspond to field sweeps up (down). Curves are offset for clarity.}
	\label{sfig-gate2}
\end{figure*}

Figure~\ref{sfig-gate2} shows UCFs measured upon repeated field sweeps after changing the top gate voltage from -4~V $\rightarrow$ 4~V over the course of several hours. Each pair of field sweeps took a total of 12~minutes to acquire. Even on such short time scales, differences between the measured conductance fluctuations are clear. While changes to UCF with charge carrier density would be interesting to study, to avoid conflating conductance changes from UCF with conductance changes from time-dependent charge trap (de-)population, all data presented elsewhere in this paper were acquired with the gate grounded and at least 12 hours after the last gate voltage change. 

\clearpage

\section{Additional device} \label{extra-data}

\begin{figure*}[h]
\centering
	\includegraphics[width=17cm]{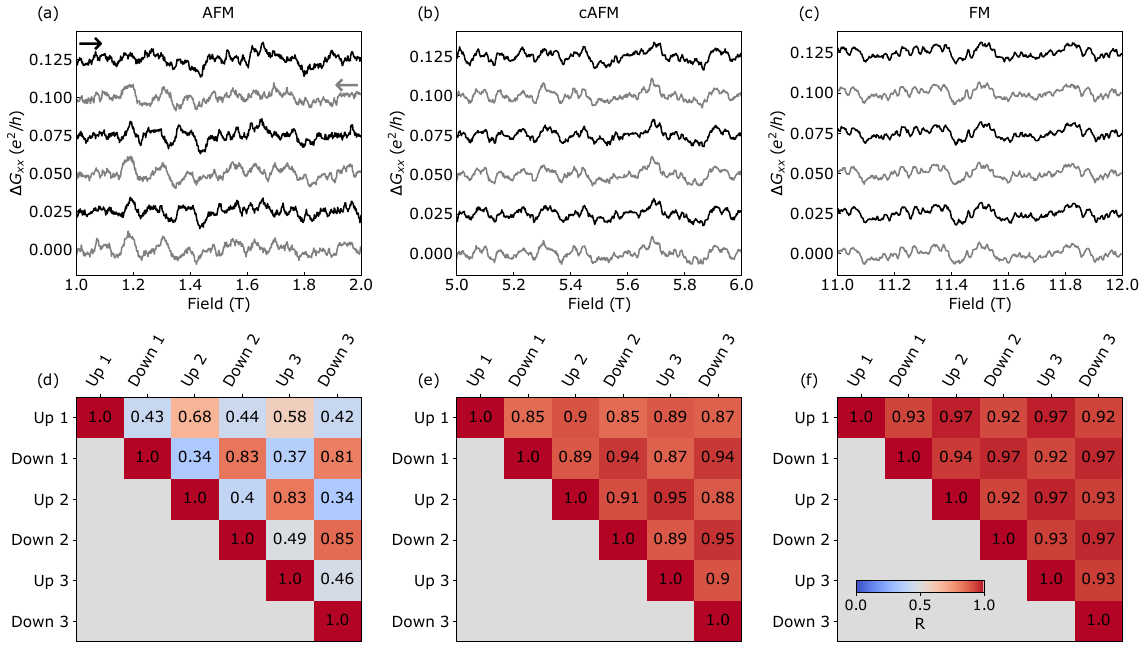}
	\caption{Reproduction of main text Figure~\ref{fig3} for data acquired on Device B's top longitudinal contact pair $V_{xxt}$. This data was acquired simultaneously to that shown in Figure~\ref{fig2}. (a-c) $\Delta G_{xx}$ as a function of applied field for sweeps up (black) and down (grey) in the (a) AFM, (b) cAFM, and (c) FM phases. For each magnetic phase, field was swept back and forth across a 1 T range with 2 mT steps in immediately subsequent runs. Data was acquired with a 2 nA ac bias current. (d-f) Correlation $R$ between each separate pair of $\Delta G_{xx}$ traces within the (d) AFM, (e) cAFM, and (f) FM phases.}
	\label{sfig-extra2}
\end{figure*}

All data presented so far was measured at the top (longitudinal) or left (Hall) voltage pairs on Device A. Simultaneously, the bottom and right voltage pairs on Device A, as well as all four different voltage pairs on Device B, were also measured. We observed no unexpected differences between these additional datasets and those presented so far; as an example, the same plots as those shown in the main text Figure~\ref{fig2} are plotted below for the additional device (Figures~\ref{sfig-extra2}). In all cases, the same dependence of correlation on field sweep direction is observed, and the overall magnitude of the correlation is similar. The one notable expected distinction between the Device A measurements and Device B is caused by geometric differences; since UCF are geometrically averaged out as the device size spans higher multiples of $l_{\phi}$, we expect a $\sqrt{4}$ reduction in the UCF magnitude in Device B. This reduction in $\Delta G_\mathrm{rms}$ is roughly observed: in the AFM phase, $\Delta G_\mathrm{rms}= 0.0037\:e^2/h$ (Device A, $0.0048\:e^2/h$); in the cAFM phase, $\Delta G_\mathrm{rms}= 0.0032\:e^2/h$ (Device A, $0.0043\:e^2/h$); in the FM phase, $\Delta G_\mathrm{rms}= 0.0030\:e^2/h$ (Device A, $0.0049\:e^2/h$).

\vspace{3ex}
\small
\noindent\textbf{\Large{Supplemental References}}\\
\noindent[S1]\hspace{2ex}L. Tai et al., \textit{ACS Nano}, 2022, \textbf{16}, 17336.\\
\noindent[S2]\hspace{2ex}Q. Jiang et al., \textit{Phys. Rev. B}, 2021, \textbf{103}, 20, 205111.\\
\noindent[S3]\hspace{2ex}C. X. Trang et al., \textit{ACS Nano}, 2021, \textbf{15}, 8, 13444--13452.\\
\noindent[S4]\hspace{2ex}Y. Aharonov and D. Bohm, \textit{Phys. Rev.}, 1959, \textbf{115}, 3, 485--491.\\

\end{document}